\documentclass[a4paper,11pt]{article}
\usepackage{pos}
\usepackage{subcaption}
\usepackage[backend=biber,style=numeric-comp,sorting=none]{biblatex}
\usepackage{slashed}

\newcommand{\be}{\begin{equation}}
\newcommand{\ee}{\end{equation}}

\newcommand{\bea}{\begin{eqnarray}}
\newcommand{\eea}{\end{eqnarray}}

\newcommand{\bei}{\begin{itemize}}
\newcommand{\eei}{\end{itemize}}

\newcommand{\nn}{\nonumber}

\newcommand\eqn[1]     {eq.\,(\ref{#1})}
\newcommand\eqns[2]    {eqs.\,(\ref{#1}) and~(\ref{#2})}

\newcommand{\T}{{\bf T}}

\newcommand{\as}{\alpha_s}
\newcommand{\eps}{\epsilon}
\newcommand{\ord}{{\cal O}}

\allowdisplaybreaks
\title{Threshold resummation of quark-gluon partonic 
channels at next-to-leading power}
\ShortTitle{Threshold resummation of quark-gluon partonic channels at NLP}

\author*[a]{Leonardo Vernazza}

\affiliation[a]{INFN, Sezione di Torino, Via P. Giuria 1, I-10125 Torino, Italy}

\emailAdd{leonardo.vernazza@to.infn.it}

\abstract{We discuss recent progress concerning the resummation 
of large logarithms at next-to-leading power (NLP) in scattering 
processes near threshold. We begin by briefly reviewing the 
diagrammatic and SCET approach, which are used to derive 
factorization theorems for physical observables in this kinematic 
limit. Then, we focus on the quark-gluon channel in deep inelastic 
scattering and Drell-Yan. We show that the use of consistency 
conditions for the cancellation of leading poles in the hadronic 
cross section can be used to achieve the resummation of large 
leading logarithms (LLs) at NLP, both within diagrammatic and 
SCET methods. In this context it is also possible to investigate 
the problem of endpoint divergences appearing at NLP in SCET, 
and relate its solution to the concept of 
re-factorization.}

\FullConference{%
  Loops and Legs in Quantum Field Theory - LL2022,\\
  25-30 April, 2022\\
  Ettal, Germany
}

\begin{document}
\maketitle

\section{Particle scattering near threshold}

\begin{figure}[t]
\centering
\includegraphics[width=0.75\textwidth]{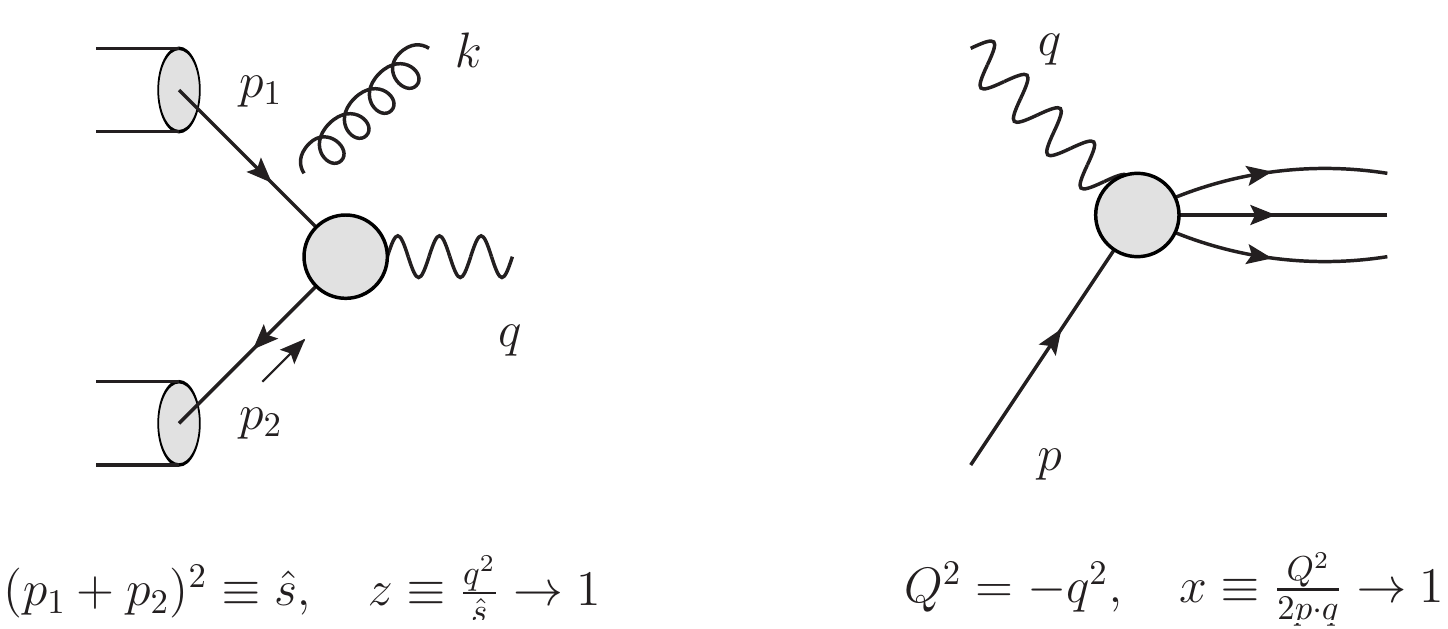}
\caption{Kinematic definition of Drell-Yan and deep inelastic scattering near threshold.}
\label{fig:figure1}
\end{figure}
In this talk we discuss scattering processes 
near threshold, focusing on Drell-Yan (DY) and 
deep inelastic scattering (DIS), shown respectively 
in the left and right diagrams of figure 
\ref{fig:figure1}. In this limit the partonic 
cross section is organized as a power expansion 
\be\label{Deltathresh}
\Delta_{ab}(\xi) 
\sim \sum_{n=0}^\infty 
\bigg(\frac{\alpha_s}{\pi}\bigg)^n\bigg[  
c_n\delta(1-\xi) 
+\sum_{m=0}^{2 n-1}\bigg(
c_{nm}\left[\frac{\ln^m(1-\xi)}{1-\xi}\right]_+
+ d_{nm} \ln^m(1-\xi) \bigg) +\ldots \bigg],
\ee
where $\xi = z$ for DY, and $\xi = x$
for DIS. The terms proportional to $c_{nm}$
and $c_n$ contribute at leading power (LP)
in $1-\xi$, while the terms $d_{nm}$
contribute at next-to-leading power (NLP). 
The large logarithms spoil the convergence 
of the perturbative expansion, and need
to be resummed. The summation of LP 
logarithms has been known for a long time,
since the seminal papers~\cite{Parisi:1979xd,Curci:1979am,Sterman:1986aj,Catani:1989ne,Catani:1990rp}. 
Later, LP threshold resummation has been 
reinterpreted and clarified using a wide 
variety of methods, including the use of 
Wilson lines~\cite{Korchemsky:1992xv,Korchemsky:1993uz}, 
the renormalization group~\cite{Forte:2002ni}, 
the connection to factorization theorems~\cite{Contopanagos:1996nh}, 
and soft collinear effective theory~\cite{Becher:2006nr,Schwartz:2007ib,Bauer:2008dt}. 
The state-of-the-art for resummation at LP is 
next-to-next-to-next-to-leading logarithmic (N$^3$LL)
accuracy for color singlet final states, and 
next-to-next-to-leading logarithmic (NNLL)
accuracy for processes involving colored 
particles in the final state.

\begin{figure}[t]
	\centering
	\includegraphics[width=0.72\textwidth]{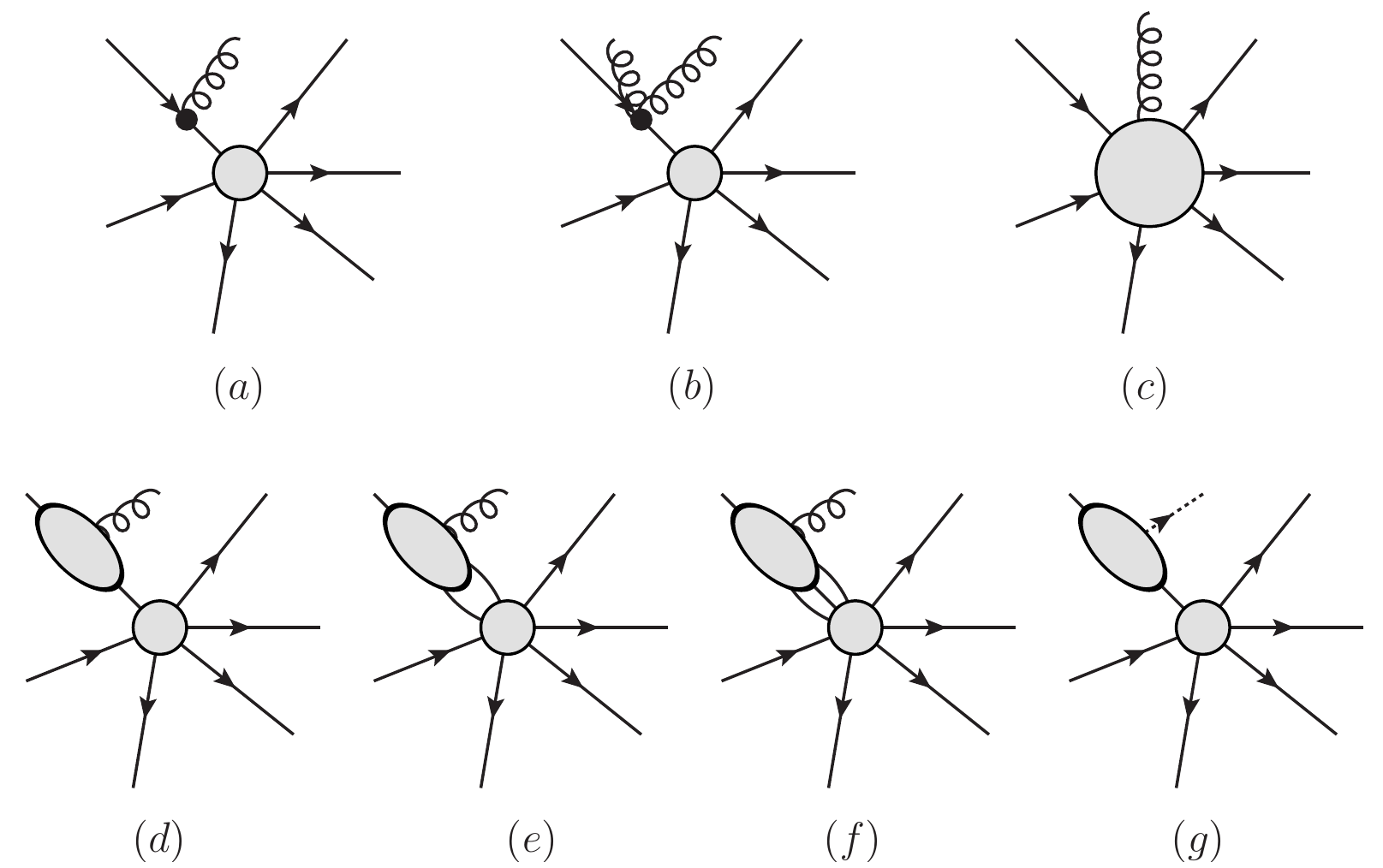}
	\caption{Diagrammatic description of soft and 
		collinear radiation contributing at NLP, 
		for scattering processes near threshold, 
		as discussed in the main text.}
	\label{fig:figure2}
\end{figure}
More recently, the quest for precision physics 
has led physicists to consider the summation of 
large logarithms at NLP, i.e. those 
multiplying the coefficients $d_{nm}$ in 
\eqn{Deltathresh}. This task is much more 
involved compared to the summation of large 
logarithms at LP and has been subject of 
intense work in the past few years, see \cite{DelDuca:1990gz,Laenen:2008ux,Laenen:2008gt,Laenen:2010uz,Bonocore:2014wua,Larkoski:2014bxa,Bonocore:2015esa,Bonocore:2016awd,Moult:2016fqy,Gervais:2017yxv,Beneke:2017ztn,Moult:2018jjd,Ebert:2018lzn,Bahjat-Abbas:2018hpv,Beneke:2018rbh,Beneke:2018gvs,Liu:2018czl,Moult:2019mog,Bahjat-Abbas:2019fqa,Beneke:2019kgv,Beneke:2019mua,Moult:2019uhz,Beneke:2019slt,Beneke:2019oqx,Liu:2019oav,Moult:2019vou,Ajjath:2020ulr,Liu:2020wbn,Anastasiou:2020vkr,Beneke:2020ibj,Ajjath:2020sjk,vanBeekveld:2021hhv,Beneke:2021aip,Bodwin:2021epw,Liu:2021mac,vanBeekveld:2021mxn,Broggio:2021fnr,Beneke:2022obx,Bell:2022ott} and references therein. 

\section{Factorization at next-to-leading power}
\label{fact}

At LP, large logarithms are related to
the emission of soft radiation, which 
is described in terms of uncorrelated 
eikonal gluons. It is easy to show that 
the uncorrelated emission of eikonal gluon 
exponentiate. Together with the 
factorization of the phase space in Mellin 
(or Laplace) space, this leads to the resummation
of large logarithms. At NLP, soft radiation 
becomes sensitive to the nature of the emitting 
particles, and begins to resolve the hard 
scattering kernels. In details, one has to take 
into account emission of soft radiation sentitive
to the spin of the emitting particle (fig.~\ref{fig:figure2} 
(a)) and multiple soft emissions (b); emission of soft 
radiation which resolves the hard interaction, 
(fig.~\ref{fig:figure2} (c)), which has been discussed 
first in \cite{Low:1958sn,Burnett:1967km}; emission 
of soft radiation from clusters of collinear virtual
particles, (fig.~\ref{fig:figure2} (d), (e), (f)).
The first of such configurations, now known as 
``radiative jets'', has been discussed in 
\cite{DelDuca:1990gz}. Last, one needs to take 
into account the emission of soft quarks, 
represented diagrammatically in 
fig.~\ref{fig:figure2} (g).

The starting point of any resummation program 
requires to be able to properly define these 
configurations in a quantum field theory 
framework. To this end, two 
main methods have been considered. One, 
referred to as diagrammatic- or direct-QCD 
approach, attempts to describe the 
configurations in figs.~\ref{fig:figure2} 
(a)--(g) in terms of matrix elements built 
of QCD fields and Wilson lines. For instance, 
the radiative jet of fig.~\ref{fig:figure2} 
(d) is defined as \cite{DelDuca:1990gz,Bonocore:2015esa,Bonocore:2016awd} 
\be
J_{\mu, a} \left( p, n, k \right) u(p) \, = \, \int d^d y \,\, 
{\rm e}^{ - {\rm i} (p - k) \cdot y} \, \left\langle 0 \left| 
\, \Phi_{n} (\infty, y) \, \psi(y) \, j_{|mu, a} (0) \, 
\right| p \right\rangle \, ,
\label{Jmudef}
\ee
where $j^\mu_a (x)$ is given by the non-abelian current
\be 
j^\mu_a (x) \, = \, g \, \left\{ - \, \overline{\psi} (x) \, \gamma^\mu \, {\bf T}_a \, \psi (x) 
\, + \, f_a^{\phantom{a} b c} \Big[ F^{\mu \nu}_c (x) \, A_{\nu \, b} (x)
+ \partial_\nu\left(A^{\mu}_b (x) A^{\nu}_c (x) \right) \Big] \right\} \, ,
\label{nabcurr}
\ee
and $\Phi_{n} (\infty, y)$ is a Wilson line 
from $\infty$ to $y$. In general, these radiative 
jets need to satisfy Ward identities. Preliminary 
results concerning the virtual collinear configurations 
in figs.~\ref{fig:figure2} (e)--(g)
have been presented in \cite{Laenen:2020nrt}
for QED, and in \cite{Gervais:2017yxv} 
for Yukawa theory.

The second method consists of an effective
field theory approach: soft and collinear modes 
in a given process are split into separate fields. 
The resulting theory is known as soft-collinear 
effective field theory (SCET)~\cite{Bauer:2000yr,Bauer:2001yt,Beneke:2002ph,Beneke:2002ni}. 
In this respect, the approach provides
a systematic power counting, such that 
the diagrammatic configurations in 
fig.~\ref{fig:figure2} are automatically
described in terms of short-distance coefficients 
times soft and collinear matrix elements, built 
from time-ordered products of SCET operators 
and power-suppressed Lagrangian 
insertions~\cite{Larkoski:2014bxa,Beneke:2017ztn,Beneke:2018rbh,Moult:2019mog}. 
For instance, within this 
approach the radiative jet of~\eqn{Jmudef}
is described in terms of a matching 
coefficient, referred to as collinear 
function~\cite{Beneke:2019oqx}: in position 
space 
\begin{eqnarray}\label{eq:exampleMatch}
i\int d^4z \,
\,\mathbf{T}\left[ \,\{\psi_{c}(tn_+)\}
\,\mathcal{L}^{(2)}_{c}(z) \right] 
=2\pi \int du \int \,dz_{-}\, \tilde{J}(t,u;z_{-}) 
\,\chi^{\text{\scriptsize PDF}}_{c}(u n_{+})\,,
\end{eqnarray}
where $\mathcal{L}^{(2)}_c$ refers to only 
the collinear pieces of the SCET Lagrangian insertion
$\mathcal{L}^{(2)}$, where the index $(2)$ denotes 
suppression by two powers of the power-counting
parameter $\lambda \sim \sqrt{1-\xi}$. In a similar 
way, one defines matching coefficients for all 
the diagrammatic structures in figs.~\ref{fig:figure2},
up to any subleading power in $\lambda$. We refer the 
reader to~\cite{Beneke:2019oqx} for further details.

\section{Endpoint divergences in SCET}

The derivation of factorization theorems 
as discussed in the previous section gives
one the tools to address the summation of 
large logarithms. Let's start by focusing
on the SCET approach, and consider for 
instance the DY invariant mass distribution
\begin{equation}
\frac{d\sigma_{\rm DY}}{dQ^2} = 
\frac{4\pi\alpha_{\rm em}^2}{3 N_c Q^4}
\sum_{a,b} \int_0^1 dx_a dx_b\,f_{a/A}(x_a)f_{b/B}(x_b)\,
(1-\eps) \, z \, \Delta_{ab}(z) \,.
\label{eq:dsigsq2}
\end{equation}
Near threshold, the leading $ab = q\bar q$
production channel receives two NLP
corrections, one of kinematic origin due to
expansion of the LP phase space, and one 
of dynamical origin, due to the power 
expansion of the partonic matrix element. 
The latter factorizes as 
\begin{eqnarray}
\label{eq:3.24}
\Delta^{dyn }_{{\rm{NLP}} }(z)&=&- \frac{2}{(1-\epsilon)} \,  
Q \left[ \left(\frac{\slashed{n}_-}{4}
\right) {\gamma}_{\perp\rho}  \left(\frac{\slashed{n}_+}{4}
\right) \gamma^{\rho}_{\perp} \right]_{\beta\gamma}
\nonumber  \\ 
&&  \hspace{0cm} \times 
\, \int d(n_+p)\,C^{A0}(n_+p, x_bn_-{p}_B  ) \,
C^{*A0}\left(\,x_an_+p_A,\,x_b{n_-p_B}\right)
\nonumber \\ 
&&  \hspace{0cm} \times \,\sum^5_{i=1}
\,\int \left\{d\omega_j\right\}
{J}_{i,\gamma\beta}\left(n_+p,x_a n_+p_A; 
\left\{\omega_j\right\} \right) \, 
{S}_{i}(\Omega; \left\{\omega_j\right\} ) 
+\rm{h.c.}\,,\quad
\end{eqnarray}
where $\Omega=Q(1-z)$. The cross section 
factorizes into the short-distance coefficients
$C^{A0}$, times the sum over terms 
containing the convolution between a set 
of collinear and soft functions. Such 
convolutions are ubiquitous in a non-local
effective field theory such as SCET: convolution
parameters represent the small component of 
collinear momenta, which have the same scaling
as the corresponding soft momentum components,
and are thus still present in the low-energy 
theory. In this respect, the convolution in 
\eqn{eq:3.24} is expected, in principle even 
at LP. The reason that one does not have to 
deal with convolutions in scattering processes 
near threshold -- at LP --  is that at this 
power order they are trival, see 
\cite{Beneke:2019oqx} for a more exhaustive 
discussion. At NLP, however, these convolutions 
become non-trivial, and present an additional 
problem: they are often divergent in $d = 4$. 
In case of DY, one of the contributions to 
\eqn{eq:3.24} explicitly reads  
\bea
\label{eq:divconvolution} 
\int_0^\Omega d\omega \,\underbrace{\big(n_+p
	\, \omega\big)^{-\epsilon}}_{{\rm{collinear\,piece}}}
\,\underbrace{\frac{1}{\omega^{1+\epsilon}}
	\frac{1}{(\Omega-\omega)^{\epsilon}}}_{{\rm{soft\,piece}}}
\,,
\eea 
which is well defined when keeping the 
exact $\eps$ dependence in the integrand, 
but diverges in $d = 4$. This divergence 
poses a problem for the standard resummation 
procedure, which relies on defining renormalized 
factors by subtracting their poles in dimensional
regularization and deriving a renormalization group 
equation for the renormalized function. Large 
logarithms are summed by evolving one function 
to the characteristic scale of the other, and 
then the convolution between the two factors 
is done. This procedure requires that the final 
convolution integral of the renormalized factors 
is well defined, which is not the case in 
\eqn{eq:divconvolution}.

\begin{figure}[t]
	\centering
	\includegraphics[width=0.75\textwidth]{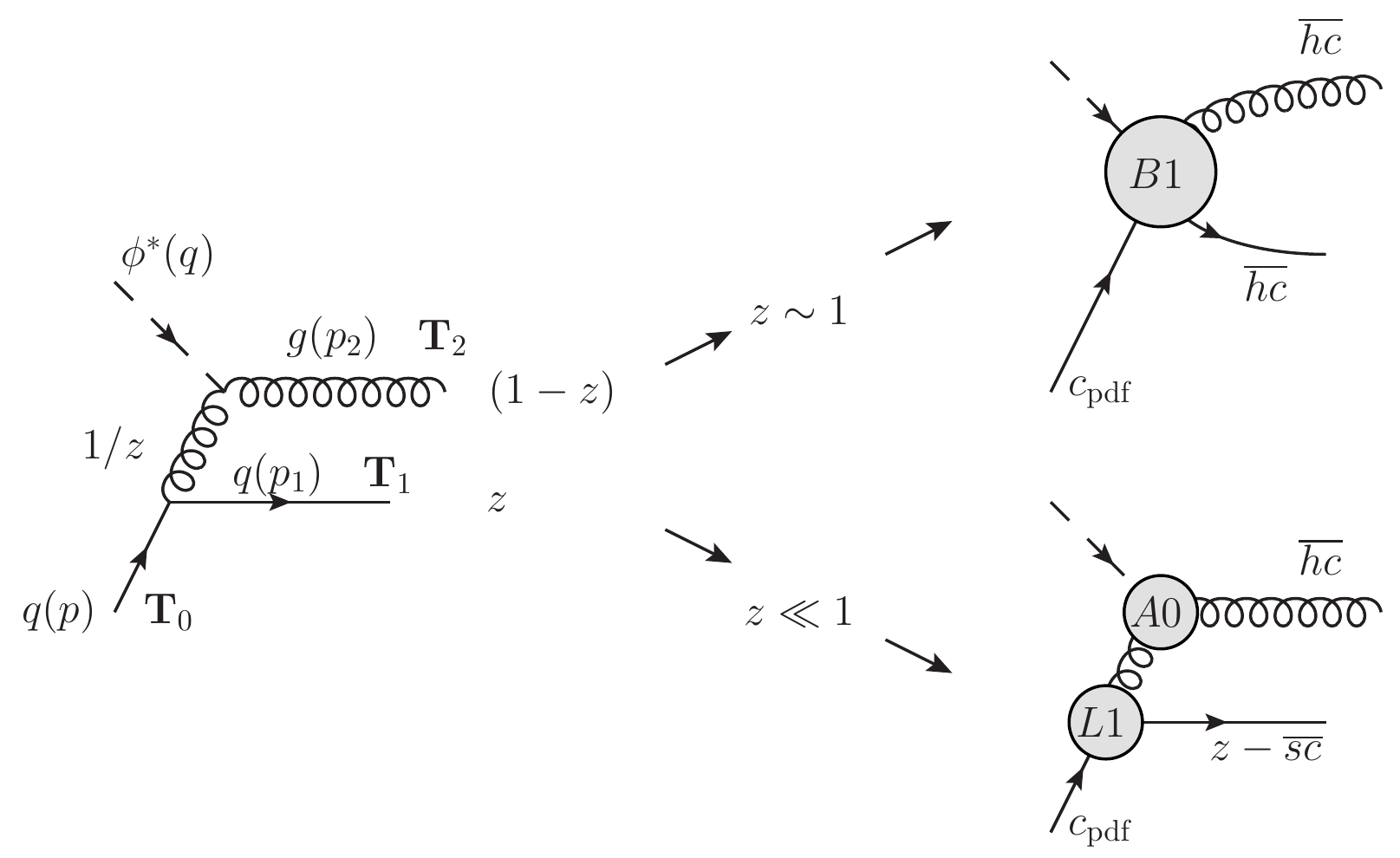}
	\caption{Left: DIS off a Higgs boson. Right: factorization 
		for generic $z \sim 1$, and for small $z \ll 1$.}
	\label{fig:figure4}
\end{figure}

To explore the problem more in detail, 
let's consider (partonic) deep inelastic 
scattering (DIS) as an example. The process is 
described in terms of structure functions, whose
factorization near threshold at LP in terms of 
parton distributions and a jet function
is well known, see \cite{Sterman:1986aj,Catani:1989ne,Korchemsky:1992xv}
and \cite{Becher:2006mr} for a SCET derivation. 
Here we focus instead on the contribution 
due to off-diagonal $qg$ channel contribution 
to DIS off a Higgs boson, see figure 
\ref{fig:figure4}. The study of this channel 
is particularly useful, because it starts at NLP. 
Therefore, compared to the Drell-Yan case discussed 
above, we have to consider only the dynamical 
contribution, i.e. the NLP matrix element, 
while the kinematic contribution (the expansion 
of the phase space in the LP matrix element) 
is absent.

The structure function for the partonic channel 
$q(p) + \phi^*(q) \to q(p_1)+g(p_2)$ is written 
as 
\be
\label{eq:TphiqVsz}
W_{\phi,q}\big|_{q\phi^*\to qg} = \int_0^1 dz\,
\left(\frac{\mu^2}{s_{qg}z\bar z}\right)^{\!\epsilon}
\mathcal{P}_{qg}(s_{qg},z)
\Big|_{s_{qg}=Q^2\frac{1-x}{x}}\, , 
\ee
where $z \equiv n_- p_1/(n_- p_1 + n_- p_2)$, 
with $\bar z=1-z$, and the momentum distribution 
function reads
\be
\label{eq:defPqg}
\mathcal{P}_{qg}(s_{qg},z) \equiv
\frac{e^{\gamma_E\epsilon}\, Q^2}{16\pi^2\Gamma(1-\epsilon)}
\frac{|{\cal  M}_{q\phi^*\to qg}|^2}{|{\cal M}_0|^2}\,,
\ee
where $|{\cal M}_0|^2$ denotes the tree-level matrix 
element squared, averaged (summed) over the spin and 
colour of the initial (final) state for the leading  
diagonal channel. At lowest order (diagram on the 
left of fig.~\ref{fig:figure4}) one has
\be\label{eq:Pqgtree}
\mathcal{P}_{qg}(s_{qg},z)\big|_{\rm tree} =
\frac{\alpha_sC_F}{2\pi}\frac{\bar z^2}{z}  + 
\mathcal{O}(\epsilon, \lambda^2) \, .
\ee
Integrating and neglecting ${\cal O}(\epsilon)$
corrections that are \emph{not} multiplied by
logarithms (i.e. counting $\epsilon\ll 1$ but
$\epsilon\ln(1-x)\sim 1$, $(1-x)^{-\epsilon}\sim 1$ 
and $\epsilon\ln(\mu/Q) \sim 1$), gives
\be
W_{\phi,q}\big|^{\rm NLP}_{\ord(\as), \, \rm leading\,pole}
= -\frac{1}{\epsilon} \,\frac{\alpha_sC_F}{2\pi}
\left(\frac{\mu^2}{Q^2(1-x)}\right)^{\!\epsilon}.
\ee 
$W_{\phi,q}\big|_{q\phi^*\to qg}$ represents the contribution
to the partonic DIS structure function when only two
partons are present in the final state. As such it is an
infrared (IR) divergent quantity. In lowest order in 
$\alpha_s$, the IR divergence is a single $1/\epsilon$ 
pole, which arises from the $z\to 0$ region of the integral
(\ref{eq:TphiqVsz}) owing to the $1/z$ behaviour of the 
tree-level momentum distribution function. 
The $z\to 0$ limit corresponds to the kinematic
configuration where the initial quark transfers all of its
momentum to the final-state gluon, and the final-state quark
becomes soft. Much like as in the DY case discussed above, it 
is therefore essential that the integration over $z$ in 
(\ref{eq:TphiqVsz}) is done in $d$ dimensions. To 
investigate further let's calculate the 1-loop correction
to the process in the left diagram of fig.~\ref{fig:figure4}.
Given that the leading order result \eqref{eq:Pqgtree} 
becomes singular only at the end point $z=0$, we can safely 
expand around this limit. Keeping only terms contributing 
to the leading poles after integration over the phase space, 
we have
\bea
\mathcal{P}_{qg}(s_{qg},z)|_{\rm 1-loop} &=&
\mathcal{P}_{qg}(s_{qg},z)|_{\rm tree}\,
\frac{\alpha_s}{\pi}\frac{1}{\epsilon^2}\,
\bigg\{{\bf T}_1\cdot {\bf T}_0\left(\frac{\mu^2}{ zQ^2}\right)^{\!\epsilon}
+ {\bf T}_2\cdot {\bf T}_0\left(\frac{\mu^2}{\bar z Q^2}\right)^{\!\epsilon} \nonumber \\
&& \hspace*{2.5cm}
+ \,{\bf T}_1\cdot {\bf T}_2\left[\left(\frac{\mu^2}{Q^2}\right)^{\!\epsilon}
- \left(\frac{\mu^2}{z Q^2}\right)^{\!\epsilon}
+\left(\frac{\mu^2}{z s_{qg}}\right)^{\!\epsilon}\,\right]
\bigg\}\,.
\label{eq:Pqg1loop}
\eea
Within SCET this result exhibits a profound problem. 
For generic $z \sim 1$ the tree amplitude in the left 
diagram of fig.~\ref{fig:figure4} corresponds to a 
$J^{B1}$ SCET operator (upper-right diagram of 
fig.~\ref{fig:figure4}) with a quark field in 
the collinear direction, and a quark and a gluon 
field in the anti-collinear direction with light-cone 
momentum fractions $z$ and $\bar{z}$, respectively. 
The tree-level matching coefficient of this operator 
is proportional to $1/z$, which gives the $1/z$ behaviour 
of $\mathcal{P}_{qg}(s_{qg},z)|_{\rm tree}$ after squaring 
the amplitude and accounting for a factor of $z$ from 
the sum of the final-state quark spin. From the 
general formula for the anomalous dimension of 
subleading power operators
\cite{Beneke:2017ztn,Beneke:2018rbh}, we get the 
double pole terms with $\T_1\cdot\T_0$ and $\T_2\cdot\T_0$ 
from the standard cusp anomalous dimension terms. However, 
one cannot obtain a cusp term for the two fields within 
the same collinear sector, i.e.~the $\T_1\cdot\T_2$ term.
In this part, there are three terms involving three 
different scales. The third, containing the scale $zs_{qg}$, 
may be disregarded here, because the dependence on 
$s_{qg}$ identifies it as a term related to the final-state 
jet function, rather than the renormalization of the 
$J^{B1}$ operator at the hard DIS vertex.
The first two terms, however, contain the hard scales 
$Q^2$ and $zQ^2$, and they are supposed to be predicted 
by the corresponding anomalous dimension. However, the 
anomalous dimension given in \cite{Beneke:2017ztn,Beneke:2018rbh} 
applies when the convolution of the coefficient function 
with the anomalous dimension is convergent, which is not 
the case here. The difference between these two terms 
is $\mathcal{O}(\epsilon)$ and hence does not contribute 
to the double pole. Instead, the expansion in $\epsilon$ 
produces $1/\epsilon\times \ln z$. However, the important 
point is that the $1/z$ singularity of the matching 
coefficient promotes this term to the same leading-pole 
order $1/\epsilon^3$ as the standard double pole terms 
after integration over $z$ as in \eqref{eq:TphiqVsz}. 
Moreover, the integral over $z$ must itself be regularized 
due to the singularity at $z=0$, and the correct result 
is obtained by {\em not} expanding (\ref{eq:Pqg1loop}) 
before integration. This can easily be seen by comparing 
(no expansion before integration)
\begin{equation}
	\frac{1}{\epsilon^2} 
	\int_0^1dz\,\frac{1}{z^{1+\epsilon}}\,(1-z^{-\epsilon})
	=-\frac{1}{2\epsilon^3}
\end{equation}
to (expansion of (\ref{eq:Pqg1loop}) before integration)
\begin{equation}
	\frac{1 }{\epsilon^2} \int_0^1dz\,\frac{1}{z^{1+\epsilon}}\, 
	\left(\epsilon\ln z -\frac{\epsilon^2}{2!}\ln^2 z+\frac{\epsilon^2}{3!}\ln^3 z+\cdots  \right)
	=-\frac{1}{\epsilon^3}+\frac{1}{\epsilon^3}-\frac{1}{\epsilon^3}
	+\cdots\,.
\end{equation}
If only the pole part of the integrand were kept, the result would 
be incomplete. This explains why it was necessary to keep the 
exact $d$-dimensional coefficient of the double pole terms in 
the one-loop momentum distribution function.

How to interpret this result? The lower-right diagram
in fig.~\ref{fig:figure4} provides the missing piece 
of this puzzle. The ultimate reason why the naive 
SCET approach fails for $z\to 0$ relies on the fact 
that, for $z \to 0$, the short-distance coefficient 
$C^{B1}$ of the corresponding operator $J^{B1}$
becomes effectively a function of two scales:
$C^{B1}(Q^2, z Q^2)$, with $z Q^2 \ll Q^2$. 
In this limit the process cannot be described 
by the EFT diagram in the upper-right diagram 
in fig.~\ref{fig:figure4}; the incoming 
PDF-collinear quark emits a $z$-anti-softcollinear
quark; the resulting propagator, proportional to 
$1/z$, is not hard, and cannot be integrated out. 
Rather, the correct EFT descriptions of the 
$z \to 0$ limit is given by the lower-right 
diagram in fig.~\ref{fig:figure4}, which 
implies the \emph{re-factorization} of the 
two-scale short-distance coefficient $C^{B1}$
according to 
\be\label{refactorization}
C^{B1}(Q,z) \, J^{B1}(z) \, \stackrel{z\to 0}{\longrightarrow} \,
C^{A0}(Q^2) \int d^4 x \, {\bf T}\Big[ J^{A0}, 
{\cal L}_{\xi q_{z-\overline{sc}}}(x) \Big] \, = \, 
C^{A0}(Q^2) \, D^{B1}(zQ^2,\mu^2) \, J^{B1}_{z-\overline{sc}}.
\ee
For $z \to 0$ the short-distance coefficient  
$C^{B1}(Q,z)$ re-factorizes into the coefficient 
$C^{A0}(Q^2)$ of the LP operator $J^{A0}$ times
a ``jet'' function $D^{B1}(zQ^2,\mu^2)$, which
at first order is given by the quark propagator 
proportional to $1/z$.

As a consequence, a correct treatment of 
off-diagonal DIS at NLP near threshold would need 
to include both the upper and lower diagrams in 
fig.~\ref{fig:figure4}, each providing the correct 
description in the corresponding $z$ regime: 
$z \sim 1$ and $z\to 0$. Crucially, both 
contributions would develop endpoint divergences. 
However, given the discussion above, it is now 
easier to understand that these are just an 
artifact of the EFT, due to splitting the matrix 
element into the two contributions above 
(the l.h.s and r.h.s of \eqn{refactorization}).
As such, the endpoint divergences in the 
two contributions are expected to cancel
each other. This allows us to conclude that 
a well defined resummation by means of 
standard EFTs methods could be achieved 
by reshuffling the endpoint divergences 
among the two contributions, such that 
both become finite. This is indeed a 
non-trivial task, which has been recently 
achieved in the context of SCET II for 
$H \to \gamma \gamma$, see 
\cite{Liu:2019oav,Liu:2020wbn},
and in the context of SCET I for 
thrust, see \cite{Beneke:2022obx}.
This work has been discussed at this 
conference, too, and we refer to
\cite{Beneke:2022zkz} for a short summary.
In this talk we illustrate another method, 
which does not exploit a full EFT treatment, 
but allows one to achieve the correct 
resummation of large logarithms and  
further clarify the non-trivial
structure of large logarithms 
near threshold in presence of
soft quarks.

\section{Consistency conditions for resummation}
\label{consistencySCET}

So far we have focused on the factorization and resummation 
of partonic structure functions in DIS, which are IR and 
ultraviolet (UV) divergent. However, partonic quantities
can be defined in terms of a renormalization prescription,
which follows from the requirement that an observable must 
be finite as $\epsilon\to 0$. To be concrete, consider the 
{\em hadronic} DIS process $p+\phi^* \to X$. From standard 
factorization theorems at leading twist in $\Lambda/Q$, 
where $\Lambda$ denotes the QCD scale, we can write the 
hadronic tensor as 
\begin{equation}
W = \sum_i W_{\phi,i} \,f_i\,,
\label{eq:leadingtwistfact}
\end{equation}
where $i$ sums over all partonic scattering channels 
and $f_i$ denotes the unrenormalized parton distribution 
function (PDF) of $i$ in the proton $p$. Thus $f_i$ contains 
dimensionally regulated UV divergences. The finite,  
$\overline{\rm MS}$ subtracted parton distributions 
and partonic cross sections are related to 
$W_{\phi,i}$, $f_i$ by
\be \label{eq:APzfactor}
\tilde f_k = Z_{ki} f_i, \qquad W_{\phi,i} = \tilde C_{\phi,k} Z_{ki}\,, 
\qquad 
{\rm such\,\, that} \qquad  
W_{\phi,i}f_i=\tilde C_{\phi,k}\tilde f_k\,.
\ee
For our purposes we need to expand
\eqn{eq:leadingtwistfact} near threshold. 
Focusing on the NLP terms one has 
\be
\sum_i(W_{\phi,i} f_i)^{\rm NLP} = 
W_{\phi,q}^{\rm NLP} f_q^{\rm LP} 
+ W_{\phi,\bar q}^{\rm NLP}f_{\bar q}^{\rm LP} 
+ W_{\phi,g}^{\rm NLP}  f_g^{\rm LP} 
+ W_{\phi,g}^{\rm LP}  f_g^{\rm NLP}\,.
\ee
We regard the PDFs in this equation as the unrenormalized 
PDFs at the factorization scale $\mu$, and to make the 
dependence on the collinear and soft-collinear scale 
explicit, we relate it to a non-perturbative reference 
PDFs via 
\bea
f_g^{\rm LP}(\mu) &=& U_{gg}^{\rm LP}(\mu)f_g(\Lambda)\,, 
\qquad 
f_q^{\rm LP}(\mu) = U_{qq}^{\rm LP}(\mu)f_q(\Lambda)\qquad
\mbox{(similarly for $\bar{q}$)}\,, \nn\\[0.2cm]
f_g^{\rm NLP}(\mu) &=& U_{gg}^{\rm NLP}(\mu)f_g(\Lambda)
+U_{gq}^{\rm NLP}(\mu)\,(f_q(\Lambda)+f_{\bar q}(\Lambda))\,.
\eea
The hadronic cross section should be finite 
for any choice of non-perturbative initial 
conditions $f_g(\Lambda)$, $f_q(\Lambda)$ and 
$f_{\bar q}(\Lambda)$. For the off-diagonal 
quark-gluon channel we focus on the terms 
proportional to $ f_q(\Lambda)$, given by 
\be 
\label{eq:Tpgifi}
\sum_i(W_{\phi,i} f_i)^{\rm NLP}\Big|_{\propto f_q(\Lambda)} = 
\left( W_{\phi,q}^{\rm NLP} U_{qq}^{\rm LP} + 
W_{\phi,g}^{\rm LP} U_{gq}^{\rm NLP}\right) f_q(\Lambda)\,.
\ee
The requirement that $W$ must be finite implies 
so-called consistency relations, which allow one
to deduce the expansion in $\epsilon$ of 
unrenormalized partonic quantities $W_{\phi,i}$ 
based on partial information. Indeed, the first 
LL resummation of the quark-gluon splitting function
was obtained in \cite{Vogt:2010cv} from the requirement 
that the DIS cross section is finite, together with 
additional assumptions on the all-order colour 
structure as well as an exponentiation ansatz
for the full partonic cross section.

Here we consider a stronger form of consistency 
relations from pole cancellations, that can be 
obtained when the regions of virtuality relevant 
to the observable are known. The different 
scaling of every region with the dimensionless 
parameters of the problem implies a larger 
number of consistency relations. In case of 
DIS near threshold, the factorization formula 
involves hard and collinear physics related 
to the scales $Q$ and $\Lambda$, which is 
non-perturbative and factorized into the PDFs. 
Near threshold the small invariant mass of the 
final state introduces a new scale into the 
problem, which is also the source of the large 
logarithms that we wish to sum. In this section 
we consider the DIS partonic cross section in 
Mellin space, according to the standard definition 
$g(N)\equiv \int_0^1 dx \,x^{N-1} g(x)$. The $x\to 1$ 
limit corresponds then to $N\to \infty$ in moment space. 
The four relevant virtualities are:
\be 
\begin{array}{ll}
\mbox{hard,}\,\,\, p^2=Q^2  \qquad &
\mbox{anti-hardcollinear,}\,\,\, p^2=Q^2\lambda^2=Q^2/N \\
\mbox{collinear,}\,\,\, p^2=\Lambda^2 \qquad  &
\mbox{softcollinear,}\,\,\, p^2=\Lambda^2\lambda^2=\Lambda^2/N
\end{array}
\ee
The anti-hardcollinear virtuality arises from the 
requirement of a small-mass final state $X$. In the 
adopted large-momentum frame, its large momentum is 
in the opposite direction of the incoming proton, 
hence ``anti-hardcollinear''. We also need a 
softcollinear virtuality $\Lambda/N\ll \Lambda$, 
which accounts for the anomalously small momentum 
of the target remnant as $x\to 1$ \cite{Becher:2006mr}. 

The calculation of the DIS process is imagined 
to be strictly factorized into contributions 
from the different virtualities. 
A multi-loop diagram is considered as a sum 
of terms, in which every loop momentum has one 
of the above virtualities, in the spirit of the 
strategy of expanding by regions \cite{Beneke:1997zp}. 
Each loop is then associated with a factor 
$(\mu^2/p^2)^\epsilon$ times a function of 
$\epsilon$, which will usually be singular.  
According to this reasoning we express the 
NLP contribution to DIS as
\bea
\label{eq:DISNLPleadingpole}
\sum_i (W_{\phi,i} f_i)^{\rm NLP} &=& 
f_q(\Lambda)\times \frac{1}{N}
\sum_{n=1} \left(\frac{\alpha_s}{4\pi}\right)^{\!n} 
\frac{1}{\epsilon^{2n-1}}\sum_{k=0}^n\sum_{j=0}^{n} c_{kj}^{(n)}(\epsilon) \left(\frac{\mu^{2n}N^j}{Q^{2k}\Lambda^{2(n-k)}}\right)^{\!\epsilon} 
\nn\\
&&+ \,f_{\bar q}(\Lambda), \,f_g(\Lambda) \,\,\,\mbox{terms}\,,
\eea
i.e., the perturbative expansion of the 
NLP contribution is expressed in terms of 
$(n+1)^2$ coefficients $c_{kj}^{(n)}(\epsilon)$ 
at order $n$. The consistency relations follow 
from the requirement that the sum of all terms 
is non-singular as $\epsilon\to 0$. In particular, 
this gives immediately 
\be
\label{eq:ckjconstraintNLP}
\sum_{k=0}^n\sum_{j=0}^{n} j^r k^{s} c_{kj}^{(n)}=0 
\qquad\mbox{for}\ s+r< 2n-1,\ r,s\geq 0 \,.
\ee
After some elaboration it is possible 
to show \cite{Beneke:2020ibj} that these 
equations allow one to determine the 
$(n+1)^2$ coefficients in terms of three 
unknown for each order $n$. However, two 
of the three “initial conditions” at every 
$n$ can be fixed trivially. In the absence 
of collinear and softcollinear loops $(k = n)$, 
there must be at least one anti-hardcollinear 
loop, since the final state cannot be made up 
of hard modes for $x\to 1$. This implies
\be
c_{n0}^{(n)}=0\,,
\ee
for all $n$. Similarly, without any hard or 
anti-hardcollinear loops ($k=0$), the necessary 
off-diagonal $q\to q g$ splitting always produces 
a softcollinear quark. Thus there must be 
at least one softcollinear loop, such that
\be
c_{00}^{(n)}=0\qquad \mbox{for all $n$\,.}
\ee
We are left with a single unknown coefficient 
at each order $n$. Hence, at each order in 
perturbation theory we can reconstruct the whole 
result by knowing the contribution of a single 
region. 

This is where the result in \eqn{eq:Pqg1loop}
comes to play. Let us recall that \eqn{eq:Pqg1loop}
gives the 1-loop virtual correction to the tree-level 
diagram in figure \ref{fig:figure4}. In this equation, 
setting $\bar z = 1$, the contribution proportional to 
the scales $(\mu^2/Q^2)^{\eps}$ and $[\mu^2/(zQ^2)]^{\eps}$
corresponds to the hard region, i.e. the coefficient 
$c_{21}^{(2)}$. As we will discuss shortly, SCET 
reasoning allows one to expect that such contribution 
exponentiate: under this assumption, \eqn{eq:Pqg1loop} 
gives 
\bea\label{eq:conjectureqghard}
\mathcal{P}_{qg, \rm hard}(s_{qg},z) &=&
\frac{\alpha_sC_F}{2\pi}\frac{1}{z} \,
\exp\bigg\{\frac{\alpha_s}{\pi}\frac{1}{\epsilon^2}\,
\bigg[(\T_2\cdot\T_0+ \T_1\cdot\T_2)
\left(\frac{\mu^2}{Q^2}\right)^{\!\epsilon} \nn \\
&& \hspace{2.0cm}
+ \,(\T_1\cdot\T_0-\T_1\cdot\T_2)
\left(\frac{\mu^2}{zQ^2}\right)^{\!\epsilon}\,
\bigg]+{\cal O}\left(\frac{1}{\epsilon}\right)\bigg\}\,,
\eea
which provides us with the whole tower of 
coefficients $c_{n1}^{(2)}$, i.e., the remaining
``initial condition'' per loop that we needed. 
This information is sufficient to determine 
the complete hadronic cross section. After some 
elaboration, we are able to determine the 
unknown terms in \eqn{eq:Tpgifi}, namely, 
$W_{\phi,q}^{\rm NLP}$ and $U_{gq}^{\rm NLP}$: 
\bea \nn
W_{\phi,q}^{\rm NLP,LL} &=& 
-\frac{1}{2N}\frac{C_F}{C_F-C_A}
\frac{\epsilon N^\epsilon}{N^\epsilon-1}
\bigg\{\exp\left[\frac{\alpha_s C_F}{\pi}
\frac{N^\epsilon-1}{\epsilon^2}
\left(\frac{\mu^2}{Q^2}\right)^\epsilon\right]
- \exp\left[\frac{\alpha_s C_A}{\pi}
\frac{N^\epsilon-1}{\epsilon^2}
\left(\frac{\mu^2}{Q^2}\right)^\epsilon\right]\bigg\},
\label{eq:TphiqNLPsol} \\ && \\ \nn 
U_{gq}^{\rm NLP,LL} &=& 
-\frac{1}{2N}\frac{C_F}{C_F-C_A}
\frac{\epsilon N^\epsilon}{N^\epsilon-1}
\bigg\{\exp\left[\frac{\alpha_s C_F}{\pi}
\frac{1-N^\epsilon}{\epsilon^2}
\left(\frac{\mu^2}{\Lambda^2}\right)^\epsilon\right]
- \exp\left[\frac{\alpha_s C_A}{\pi}
\frac{1-N^\epsilon}{\epsilon^2}
\left(\frac{\mu^2}{\Lambda^2}\right)^\epsilon\right]\bigg\}.
\\
\label{eq:UphiqNLPsol}
\eea
From here it is also possible to obtain 
the corresponding renormalized coefficients,
according to \eqn{eq:APzfactor}, 
and we refer to \cite{Beneke:2020ibj} for a 
throughout derivation.  

It remains to justify the exponentiation 
hypothesis in \eqn{eq:conjectureqghard}. 
Indeed, even without a proper treatment 
of endpoint divergences, the exponentiation 
of the 1-loop hard region can be explained 
within the refactorization condition of 
\eqn{refactorization}. Focusing on the r.h.s. 
of \eqn{refactorization}, and working in 
$d = 4-2\eps$, the evolution of the coefficient 
$C^{B1}(Q,z)$ can be obtained as a two-step 
procedure, in which the LP short-distance 
coefficient $C^{A0}$ and the jet $D^{B1}$ are 
evolved to a common scale $\mu$. The anomalous 
dimension of $C^{A0}$ is well known, and the 
one of $D^{B1}$ can be determined by means of
a region calculation, obtaining
\begin{align}
\left[C^{A0}\left(Q^{2},\mu^{2}\right)\right]_{\rm bare}&=C^{A0}\left(Q^{2},Q^{2}\right)\exp\left[-\frac{\alpha_s C_A}{2\pi}\frac{1}{\epsilon^{2}}
\left(\frac{Q^{2}}{\mu^{2}}\right)^{\!-\epsilon}\right],\nn \\
\left[D^{B1}\left(zQ^{2},\mu^{2}\right)\right]_{\rm bare}
&=D^{B1}\left(zQ^{2},zQ^{2}\right)
\exp\left[-\frac{\alpha_s}{2\pi}\left(C_{F}-C_{A}\right)
\frac{1}{\epsilon^{2}}\left(\frac{zQ^{2}}{\mu^{2}}\right)^{\!-\epsilon}\right].
\label{eq:CDresum}
\end{align}
Replacing the appropriate values 
$\T_1\cdot\T_0=C_A/2-C_F$, $\T_2\cdot\T_0=\T_1\cdot\T_2=-C_A/2$
in \eqn{eq:conjectureqghard}, we see that the evolution of the 
r.h.s of \eqn{refactorization} according to \eqn{eq:CDresum}
reproduces \eqn{eq:conjectureqghard}, thus providing a 
SCET-based justification for the ``soft quark Sudakov''
exponentiation conjecture.
 
Let us conclude this section with an interesting 
observation: it is remarkable that, in the leading-pole 
approximation, the full result for $W_{\phi,q}^{\rm NLP,LL}$ 
in \eqn{eq:TphiqNLPsol} follows from the exponentiation 
conjecture for the hard-only amplitude, \eqn{eq:conjectureqghard}, 
by a simple substitution. Let us define 
\be \label{ASdef}
A \equiv \frac{\alpha_s (C_F-C_A)}{\pi} \frac{1}{\epsilon^2} 
\left(\frac{\mu^2}{Q^2}\right)^{\!\epsilon}, 
\qquad 
S \equiv  \frac{\alpha_s C_A}{\pi} \frac{1}{\epsilon^2} 
\left(\frac{\mu^2}{Q^2}\right)^{\!\epsilon}\,.
\ee
Then, integrating the amplitude in  
\eqn{eq:conjectureqghard} against the phase 
space in \eqn{eq:TphiqVsz} gives the hard-only 
contribution to the structure function, which 
reads
\be
\label{eq:NLPsolsubstitute}
W_{\phi,q}\Big|_{q\phi^*\to qg}^{\rm hard}=  
\frac{1}{N}\frac{\alpha_sC_F}{2\pi\epsilon}
\left(\frac{\mu^2N}{Q^2}\right)^{\!\epsilon}
\exp\left[-S\right]
\,\times  \frac{\exp(-A)-1}{A}.
\ee
In the same notation, the full structure 
function of \eqn{eq:TphiqNLPsol} reads 
\be\label{eq:TphiqNLPsolB}
W_{\phi,q}^{\rm NLP,LL} = 
-\frac{1}{N}\frac{\alpha_sC_F}{2\pi\epsilon}
\left(\frac{\mu^2N}{Q^2}\right)^{\!\epsilon}
\exp\left[S \,(N^\epsilon-1)\right]
\,\times \,\frac{\exp(A \,(N^\epsilon-1))-1}{A\,(N^\epsilon-1)}\,,
\ee
i.e., the full structure function can be obtained from the 
result in the hard region by replacing $A\to A \,(1-N^\epsilon)$, 
$S\to S \,(1-N^\epsilon)$. The appearance of the factor 
$(N^\epsilon-1)$ is characteristic of the leading-pole 
solution.

\section{Resummation within a diagrammatic approach}

In section \ref{fact} we have discussed also another 
approach, based on diagrammatic methods in QCD. 
Within this framework one defines the collinear 
and soft matrix elements in fig.~\ref{fig:figure2} 
in terms of fields and Wilson lines in QCD. It is then 
possible to study the exponentiation of logarithmic 
contributions by means of a diagrammatic analysis, 
determining which diagrams contribute to a given 
power- and logarithmic-accuracy, and investigating 
their combinatorial structure (see e.g. \cite{Laenen:2010uz}). 
This procedure is typically performed in dimensional 
regularization, such that one should not have 
to deal with endpoint divergences at any stage 
of the computation. The quark-gluon channel 
in DIS provides an interesting example, 
where we can compare the SCET approach 
discussed above with the corresponding 
calculation based on diagrammatic methods. 

\begin{figure}
\centering
\includegraphics[width=0.6\textwidth]{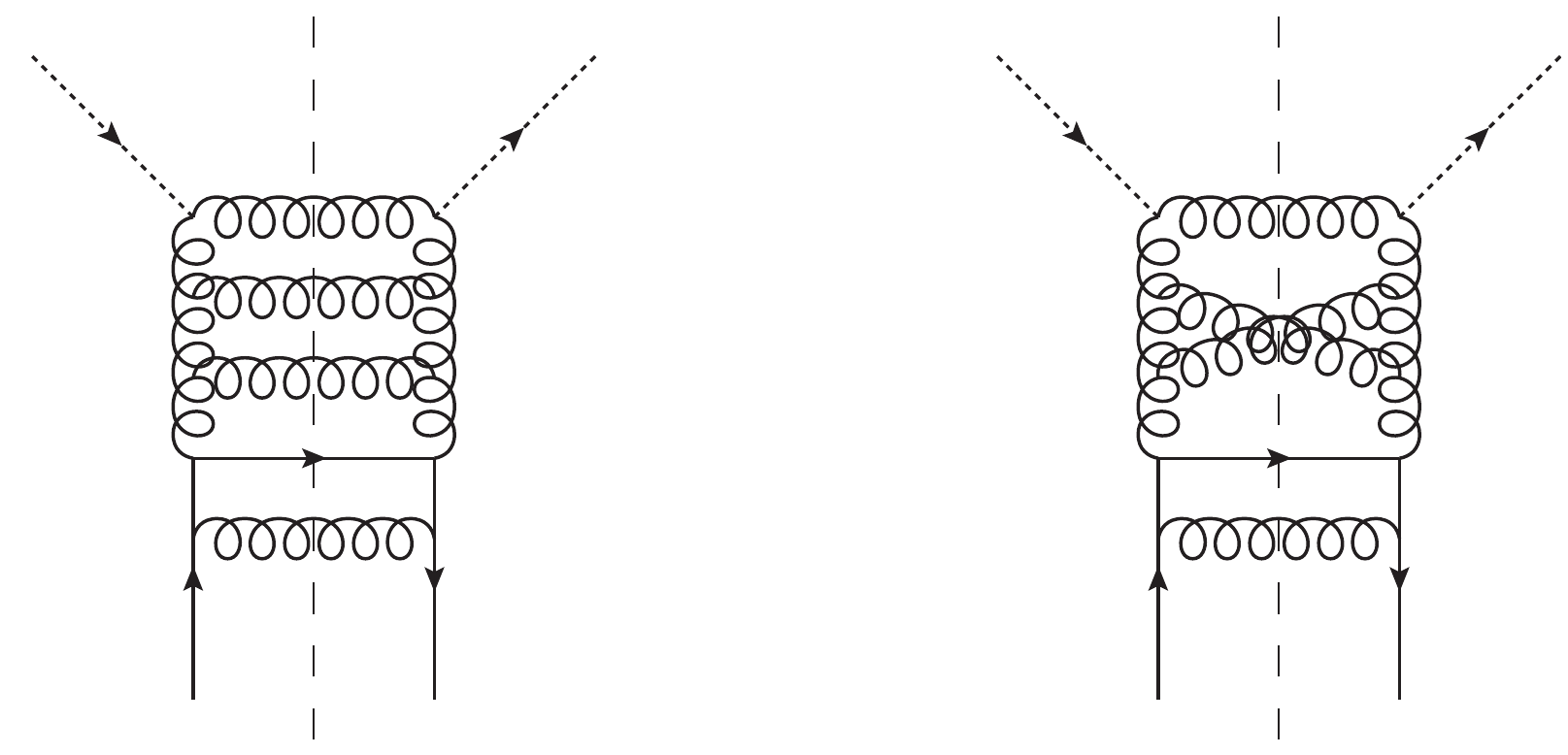}
\caption{Left: a ladder graph contributing to the DIS $qg$ channel 
at NLP LL; right: a crossed-ladder graph.}
\label{fig:ladder}
\end{figure}
The starting point for the diagrammatic analysis
\cite{vanBeekveld:2021mxn} also begins from the 
consistency relations discussed in the previous 
section. As we concluded there, the whole DIS 
cross section at NLP can be reconstructed once 
a single ``initial condition'' per loop in a 
given momentum region is known. Within the SCET 
approach it is convenient to consider the hard 
region, because the latter can be determined 
to all orders within the refactorization
procedure, as discussed in the previous section. 
Within a diagrammatic analysis, instead, it proves 
convenient to consider the soft-anticollinear region. 
Inspecting the tree-level diagram on the left in 
figure \ref{fig:figure4}, it is easy to realize 
that the quark-gluon-quark interaction vertex, 
where the initial collinear quark is converted 
into a collinear gluon by means of a soft-anticollinear
quark emission, provides a power suppression by a factor 
of $\lambda \sim \sqrt{1-x}$. Two such vertices are 
present at the level of the matrix element squared, 
thus providing the required suppression to NLP, or 
$\lambda^2 \sim (1-x)$. Any other emission in the 
soft-anticollinear region must therefore be given 
in terms of soft gluon emissions at LP, for which 
the eikonal approximation can be used. The diagram 
contributing can be further reduced by noticing that 
a) leading logarithms only arise from the kinematic 
region in which the transverse momenta of the emitted 
partons are strongly ordered; and b) one may reduce 
the set of relevant Feynman diagrams for the 
squared matrix element to those having a pure 
ladder form, as shown in the left diagram of 
fig.~\ref{fig:ladder}. Crossed ladders, such 
as the graph on the right in fig.~\ref{fig:ladder}, 
do not contribute at LL. In non-abelian theories 
such as QCD, this property is not guaranteed in 
general gauges, but can be made manifest by 
choosing to define the polarization states 
of the emitted gluons in a particular way. 
Upon choosing a reference vector $c_\mu$, 
one may define physical gluon polarization 
vectors $\epsilon_\mu(k)$ via the simultaneous 
requirements
\begin{equation}
	k\cdot \epsilon(k)=c\cdot\epsilon(k)=0\,.
	\label{physpols}
\end{equation}
If in addition $c$ is a null vector ($c^2=0$), 
the sum over physical gluon polarization states 
has the form
\begin{equation}
	\sum_{{\rm pols.}}\epsilon_\mu^\dag (k)\epsilon_\nu(k)=
	-\eta_{\mu\nu}+\frac{k_\mu c_\nu + k_\nu c_\mu}{c\cdot k}\,.
	\label{polsum}
\end{equation}
As explained in detail in refs.~\cite{Dokshitzer:1978hw,Dokshitzer:1991wu}, 
the kinematic dominance of uncrossed gluon ladders 
occurs for the explicit choice
\begin{equation}
	c=q' \equiv q+x p\,,
	\label{c=q'}
\end{equation}
\begin{figure}[t]
	\centering	
	\includegraphics[width=0.80\textwidth]{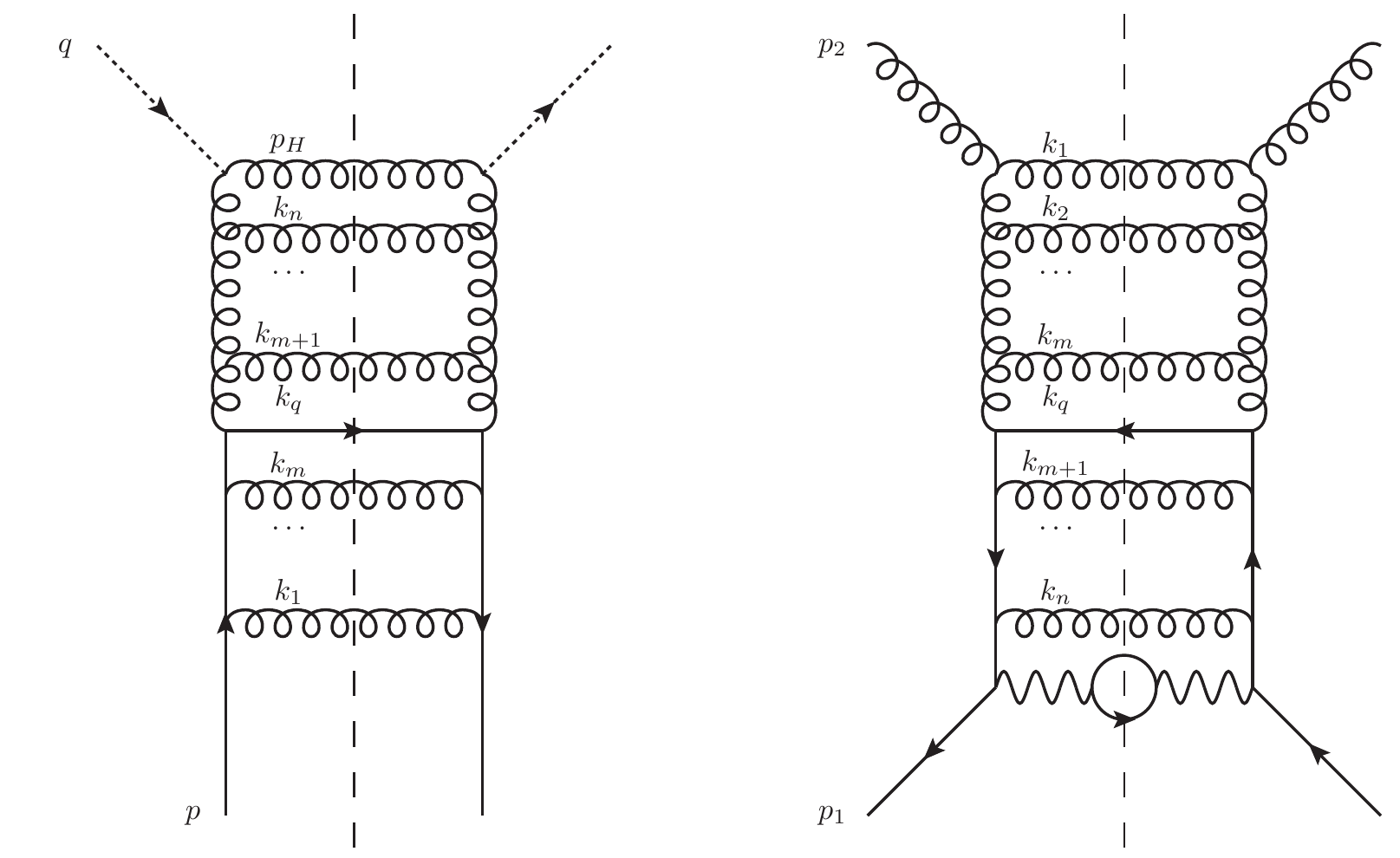}
	\caption{Left: ladder diagram contributing at LL 
		accuracy to the $qg$ channel of Higgs-induced DIS, 
		and right: ladder diagram contributing to the $g\bar q$ 
		channel in Drell-Yan.}
	\label{fig:figure5}
\end{figure}
where $q$ and $p$ are defined as in 
fig.~\ref{fig:ladder}, and $x = Q^2/(2 p\cdot q)$
is the Bjorken $x$. This reasoning allows us to
conclude that the LL contribution to the $qg$ 
channel in Higgs-induced DIS at order $n+1$
is given by the left ladder diagram in 
fig.~\ref{fig:figure5}, which corresponds 
to the amplitude 
\begin{align}
	\overline{|{\cal M}_{qh\rightarrow qg_1\dots g_n}|^2} \notag &  \\
	&\hspace{-2.0cm}
	=\,\frac{|\phi_h|^2 C_F^{m+1}C_A^{n-m}
		g_s^{2(n+1)}}{8\mu^{(d-4)(n+1)}}\left(\prod_{i=1}^n\frac{2q \cdot p\,p\cdot k_i}
	{q'\cdot k_i}\right){\rm Tr}[\slashed{p}\gamma^\beta\slashed{k}_q\gamma^\alpha]
	\left(-\eta_{\alpha\beta}+\frac{q'_\alpha p_{H,\beta}+q'_{\beta}p_{H,\alpha}}
	{q'\cdot p_H}\right)\notag\\
	&\hspace{-1.8cm}
	\times \frac{1}{(p\cdot k_1)^2[p\cdot(k_1+k_2)]^2\ldots
		[p\cdot(k_1+\ldots+k_m+k_q)]^2\ldots
		[p\cdot(k_1+\ldots +k_n+k_q)]^2},
	\label{gluoncalc1}
\end{align}
where we highlight the eikonal structure of 
the $n$-soft gluon emissions, and $\phi_h$ 
represents the effective Higgs-gluon coupling. 
The calculation of the full structure function 
requires us to integrate the matrix element 
squared above against the $(n+2)$-phase space. 
Given that the matrix element squared is already 
contributing at NLP, the phase space can be 
approximated to LP, which entails a significant 
simplification: the phase space factorizes into 
single-gluon phase spaces in Laplace space. 
After some elaboration we get the LL 
contribution at order $n+1$ (with $\mu^2 = Q^2$):
\be \label{Tphiqresult}
W_{\phi,q}^{(n+1)}=-\left(\sum_{m=0}^{n}
C_F^{m+1}C_A^{n-m}\right)\frac{2}{\epsilon}\frac{N^\epsilon}{N}
\left(\frac{4 N^\epsilon}{\epsilon^2}\right)^n \frac{1}{(n+1)!}\,,
\ee
which can be resummed into a closed form:
\be \label{Tphiqresult2} 
W^{\rm soft}_{\phi,q}\Big|_{\rm LL}
= \sum_{n=1}^\infty \bigg(\frac{\as}{4\pi}\bigg)^n W_{\phi,q}^{(n)}
= -\frac{1}{2N}\frac{C_F}{C_F-C_A}
\frac{\epsilon N^\epsilon}{N^\epsilon-1}
\bigg\{\exp\left[\frac{\alpha_s C_F}{\pi}\frac{N^\epsilon}{\epsilon^2}\right]
-\exp\left[\frac{\alpha_s C_A}{\pi}\frac{N^\epsilon}{\epsilon^2}\right]\bigg\}.
\ee
The full amplitude can be easily reconstructed 
by exploiting the consistency relations. As 
expected, we recover \eqn{eq:TphiqNLPsol}.
At this point it is interesting to notice that, 
writing \eqn{Tphiqresult2} as
\be
\label{eq:NLPsolsubstituteSoft}
W^{\rm soft}_{\phi,q}\Big|_{\rm LL} =  
-\frac{1}{N}\frac{\alpha_sC_F}{2\pi\epsilon}
\left(\frac{\mu^2N}{Q^2}\right)^{\!\epsilon}
\exp\left[S N^\epsilon\right]
\,\times  \frac{\exp(A N^\epsilon)-1}{A N^\epsilon},
\ee
where the functions $A$ and $S$ have been defined 
in \eqn{ASdef}, the calculation of the soft region 
gives the full result by means of the substitution 
$A N^\epsilon \to A (N^\epsilon-1)$, 
$S N^\epsilon \to S (N^\epsilon-1)$, which is 
indeed consistent with the observation made 
around \eqns{eq:NLPsolsubstitute}{eq:TphiqNLPsolB}.

Let us conclude this section by mentioning that the methods
of section \ref{consistencySCET} and the diagrammatic methods 
discussed here can be applied to the calculation of other 
off-diagonal processes, such as the quark-gluon channel 
contribution to DY, $g(p_1)\bar{q}(p_2) \rightarrow  
\gamma^*(q)\rightarrow e^+(q_1) e^-(q_2)$. 
Focusing on the diagrammatic method, in this case it 
is possible to show \cite{vanBeekveld:2021mxn} that 
one has to consider the tower of ladders in the right 
diagram of fig.~\ref{fig:ladder}. This gives once again 
the soft contribution to the $qg$ channel in DY, and then
the full result can be reconstructed by exploiting 
consistency relations, obtaining the resummed 
(bare) partonic cross section 
\bea \label{WDYLLcomplete} \nn
{W}^{\rm NLP, LL}_{{\rm DY},g\bar{q}} &=&
-\frac{T_R}{2(C_F-C_A)}\frac{1}{N} \frac{\epsilon(N^{\epsilon-1})}
{N^\epsilon-1}
\exp\left[\frac{4a_s C_F(N^\epsilon-1)}{\epsilon^2}\right] \\
&&\hspace{2.0cm} \times\, 
\left\{\exp\left[\frac{4a_s C_F N^\epsilon(N^\epsilon-1)}
{\epsilon^2}\right]
-\exp\left[\frac{4a_s C_A N^\epsilon(N^\epsilon-1)}
{\epsilon^2}\right]
\right\}, 
\eea
and after some work, for which we refer to 
\cite{vanBeekveld:2021mxn}, one finds the 
renormalized partonic cross section 
\be \label{CDYgqres}
\widetilde{C}_{{\rm DY},g\bar{q}}\Big|_{\rm LL}=\frac{T_R}{C_A-C_F}
\frac{1}{2N \ln N}\left[e^{8C_Fa_s\ln^2 N}{\cal B}_0[4a_s(C_A-C_F)\ln^2 N]
-e^{(2C_F+6C_A)a_s\ln^2N}\right],
\ee
where ${\cal B}_0(x)=\sum_{n=0}^\infty \frac{B_n}{(n!)^2}x^n$,
and $B_n$ are Bernoulli numbers. This result reproduces 
an earlier conjecture in ref.~\cite{Presti:2014lqa}, and 
can be obtained also within the methods discussed in 
section \ref{consistencySCET}, \cite{Beneke:2020ibjB}.

\section{Conclusions} 
\label{sec:conclusions}

The factorization of scattering processes near 
threshold beyond leading power is nontrivial  
-- it involves configurations of collinear and 
soft momenta, that can be described in terms of 
matrix elements obtained within a diagrammatic
\cite{Bonocore:2016awd,Laenen:2020nrt} or a SCET-based 
\cite{Larkoski:2014bxa,Moult:2019mog,Beneke:2019oqx} 
approach. Within the latter, traditional resummation 
methods fail \cite{Beneke:2018gvs,Beneke:2019oqx}, 
due to the appearance of endpoint divergences in 
the convolution between short-distance coefficients, 
collinear and soft functions. 

This problem can be studied conveniently in off-diagonal 
channels of $2\to 1$ or $1\to 2$ processes, such as the 
quark-gluon channel in DIS and DY, which starts at NLP, 
and where endpoint divergences appear already in the 
factorization of the partonic cross sections at LL accuracy. 

The requirement of pole cancellation in the hadronic 
cross section can be used to obtain consistency 
conditions. These can then be used together with 
the assumption of exponentiation of the 1-loop hard-region 
contribution to the partonic cross section, to achieve 
the resummation of NLP LL logarithms \cite{Beneke:2020ibj}. 
Within SCET, the exponentiation hypothesis can 
be justified in the context of a re-factorization 
involving the short-distance coefficient representing 
the hard-region contribution, which has been exploited 
in \cite{Liu:2019oav,Liu:2020wbn,Beneke:2022obx}
to construct factorization theorems free of endpoint 
divergences.

Consistency conditions can be used as well in combination
with diagrammatic methods \cite{vanBeekveld:2021mxn}. In 
this case it is possible to show that the soft region 
contribution to the partonic cross section can be 
determined and resummed to all orders in terms of 
ladder diagrams. The full cross section can then
be reconstructed by exploiting the consistency conditions.

The resummation of LLs at NLP in off-diagonal channels, 
together with the resummation of LLs in diagonal channels 
previously obtained in \cite{Moult:2018jjd,Beneke:2018gvs,Beneke:2019mua,Bahjat-Abbas:2019fqa}
completes the resummation of LLs at NLP in $2\to 1$ and $1\to 2$
processes. The methods discussed in this talk 
will be useful to extend the resummation of 
NLP large logarithms near threshold at NLL 
accuracy and beyond.

\subsubsection*{Acknowledgments} 
This work has been supported by Fellini -- Fellowship for
Innovation at INFN, funded by the European Union's Horizon
2020 research programme under the Marie Sk\l{}odowska-Curie
Cofund Action, grant agreement no.~754496.

\printbibliography

\end{document}